\documentclass[useAMS,usenatbib]{mn2e}
\usepackage[dvips]{epsfig}
\include{psfig}
\newcommand{\be}{\begin{equation}}   
\newcommand{\ee}{\end{equation}}   
\newcommand{\bea}{\begin{eqnarray}}   
\newcommand{\eea}{\end{eqnarray}}

\title[Cubic anomalies in WMAP]{Cubic anomalies in WMAP }
\author[Kate Land and Jo\~{a}o Magueijo]{
Kate Land and Jo\~{a}o Magueijo 
\thanks{E-mail:kate.land@imperial.ac.uk,j.magueijo@ic.ac.uk}
\\\\
Theoretical Physics Group, Imperial College, Prince Consort Road, 
London SW7 2BZ, UK
}
\begin{document}

\date{\today}

\pagerange{\pageref{firstpage}--\pageref{lastpage}} \pubyear{2002}

\maketitle

\label{firstpage}

\begin{abstract}
We perform a frequentist analysis of the bispectrum of WMAP first year
data. We find clear signal domination up to $\ell\approx 200$,
with overall consistency with Gaussianity, except for the following 
features. There is a flat patch (i.e. a low $\chi^2$ region)
in the same-$\ell$ components of the bispectrum spanning the range
$\ell=32-62$; this {\it may be} interpreted as ruling out Gaussianity
at the $99.6$\%  confidence level. There is also an asymmetry between
the North and South inter-$\ell$ bispectrum components 
at the $99\%$ confidence level. The preferred asymmetry axis 
correlates well with the $(\ell ,b)=(57,10)$
direction quoted in the literature for asymmetries in the power
spectrum and three-point correlation function. However our
analysis of the quadrupole (its bispectrum and principal axes)
fails to make contact with previously
claimed anomalies.
\end{abstract}


\section{Introduction}
The hypothesis of statistical Gaussianity
plays a central role in the standard theory of structure
formation. Unsurprisingly, a large amount of effort has gone
into checking this assumption, with the cosmic microwave 
background (CMB) anisotropy being its cleanest probe.
The early work on COBE-DMR data was punctuated by the first 
findings that we might not live in the best of all possible worlds. 
The data passed standard, but not very discriminative 
Gaussianity tests~\citep{kog96a}. Upon closer inspection, however,
anomalies were discovered~\citep{fmg,nfs}. For example,
the bispectrum analysis of the 4-year dataset revealed 
evidence against Gaussianity~\citep{fmg,joao}
to a high significance  level.

The possibility remained that these non-Gaussianity ``detections'' might be 
due to undocumented systematic errors. 
This hope was partly satisfied by the discovery 
of subtle errors in the map-making procedure of 
COBE-DMR~\citep{banday}. Pixelization artifacts and further systematic
errors in COBE were later found, with the aid of the higher quality
WMAP data~\citep{jooes}. Indeed all the early work was 
plagued by high noise levels, poor resolution and the threat 
of multiple systematics. The WMAP project\citep{wmap} has  opened up
the doors for the first convincing evaluation of the 
hypothesis of Gaussianity.

Ironically the WMAP data has renewed concerns over the Gaussianity
of the CMB fluctuations~\citep{copi,erik2,coles,cruz,hansen,hbg,
vielva,Komatsu2,park,lw,muk,erik1,copi1}.
Most worryingly, it appears that the data 
displays a North-South asymmetry in the power spectrum and
three-point correlation function. This questions both
the statistical isotropy and Gaussianity of the data. The
spread in power spectrum and 3-point function is sensitive to the fourth
and sixth momenta of the distribution.

In this paper we perform a frequentist analysis of the WMAP
bispectrum. This is to be contrasted with the work of~\cite{ngwmap},
where the WMAP bispectrum is analysed with reference to 
a particular non-Gaussian model,
parameterised by the non-linear coupling parameter $f_{NL}$.
Such work neglects bispectrum non-Gaussianity that might
arise in any other theory; or indeed that might be due to spurious
systematics. Our exploratory approach is to be seen as complementary 
to that by~\cite{ngwmap}. In particular we shall be concerned with
correlating our bispectrum studies and the WMAP anomalies
found by other groups, namely claims for North-South
irregularities.

Our paper is organised as follows. In Section~\ref{intbisp}
we establish our notation and bispectrum definitions.
We introduce same-$\ell$ 
and inter-$\ell$ normalised bispectrum measures,
denoted by $I_\ell$ and $J_\ell$ respectively. These
are sensitive to phase correlations within a single multipole
and between adjacent multipoles. Their normalisation ensures
that for  a Gaussian process they are statistically independent
from the power spectrum~\citep{conf}.

In Section~\ref{flat} we then evaluate $I_\ell$ and $J_\ell$
for WMAP and compare results with Monte-Carlo simulations
in which a Gaussian signal is subject to the same noise and
beam characteristics as the real data. We also discuss various
galactic masks.
We find a low $\chi^2$ region in the $I_\ell$ in the range
$\ell=32-62$. Depending on the prior, this  may be interpreted as ruling out 
Gaussianity at the $99.6$\%  confidence level (we relegate
to Section~\ref{select} a discussion of possible selection effects).

In Section~\ref{asym} we re-examine claims for North-South
asymmetries in CMB from the point of view of the bispectrum.
We find asymmetries in the inter-$\ell$ components of the 
bispectrum  which are not only statistically significant
($99\%$ confidence level, {\it even when all possible
selection effects are taken into account}),
but also correlates well with the $(\ell ,b)=(57,10)$
direction quoted in the literature. Our detection is independent 
from, but reinforces similar anomalies found
for the 3-point correlation function~\citep{erik1}.

However,  our analysis of the quadrupole 
fails to make contact with previously claimed weird features.
In Section~\ref{quadtxt} we provide a geometrical interpretation
of the quadrupole where the bispectrum may be seen as a shape
factor. In addition we identify three orthogonal principal 
directions. None of these show anything abnormal or correlate
with suspicious directions.

In Section~\ref{con} we conclude with a summary of our findings.

\section{The bispectrum}\label{intbisp}
We start by reviewing some results and
definitions pertaining to the bispectrum. Given a full-sky map,
$\frac{\Delta T}{T}({\bf n})$, this may be expanded into Spherical
Harmonic functions:
\begin{eqnarray}
\frac{\Delta T}{T}({\bf n})=\sum_\ell \delta T_\ell=
\sum_{\ell m}a_{\ell m}Y_{\ell m}({\bf n})
\label{almdef}
\end{eqnarray}
Each multipole component $\delta T_\ell$ contains fluctuations
with a characteristic angular scale inversely proportional to $\ell$.
The coefficients $a_{\ell m}$ are not rotationally invariant
but may be combined into
rotationally invariant multilinear forms (see \cite{santa}
for a possible algorithm). The only {\it bilinear} invariant is the commonly 
used angular power spectrum, given by
${\hat C}_\ell=\frac{1}{2\ell+1}\sum_m|a_{\ell m}|^2$.
This gives a measure of the overall intensity
of each multipole.
The most general
{\it cubic} invariant is the bispectrum,  and is given by
\begin{eqnarray}
{\hat B}_{\ell_1\ell_2\ell_3}&=&\frac{\left (
\begin{array}{ccc} \ell_1
 & \ell_2 & \ell_3 \\ 0 & 0 & 0
\end{array} \right )^{-1}}
{(2\ell_1+1)^{\frac{1}{2}}
(2\ell_2+1)^{\frac{1}{2}}(2\ell_3+1)^{\frac{1}{2}}}\times \\
&&\sum_{m_1m_2m_3}\left
( \begin{array}{ccc} \ell_1 & \ell_2 & \ell_3 \\ m_1 & m_2 & m_3
\end{array} \right ) a_{\ell_1 m_1}a_{\ell_2 m_2} a_{\ell_3 m_3}\nonumber
\end{eqnarray}
where the $(\ldots)$ is the Wigner $3J$ symbol.
The proportionality constant is chosen 
so as to enforce a roughly constant cosmic variance. 
There are various alternative 
ways in which the bispectrum may be computed; for instance
\be
{\hat B}_{\ell_1\ell_2\ell_3}\propto\int d\Omega \delta T_{\ell_1}
\delta T_{\ell_2}\delta T_{\ell_3}
\ee
with similar constructions for higher order moments~\citep{sg98,Komatsu}.
This may be computationally advantageous.

Selection rules require that $\ell_1+\ell_2+\ell_3$ be even, and we
 impose $\ell_{\it i}\geq$ 2.
So for even $\ell$ the choice $\ell_1=\ell_2=\ell_3=\ell$ leads to
the ``single-$\ell$'' bispectrum ${\hat B_\ell}=B_{\ell\, \ell \,\ell}$
\citep{fmg}.
Other bispectrum components are sensitive to correlations between 
different scales. The simplest chain of correlators is therefore
${\hat A_\ell}=B_{\ell-1\, \ell \,\ell+1}$ -- the ``inter-$\ell$''
bispectrum \citep{joao}.
Other components, involving more distant multipoles, 
may be considered \citep{haav}
but they are very likely to be dominated by noise; it has been observed that 
 non-Gaussian inter-scale
correlations decay with $\ell$ separation.

In this paper we shall consider ratios
\begin{eqnarray}\label{i3}
I^3_\ell &=& { {\hat B}_{\ell}
\over ({\hat C}_{\ell})^{3/2}}
 \label{defI}
\end{eqnarray}
and
\begin{eqnarray}\label{j3}
J^3_\ell &=& { {\hat A}_{\ell}
\over ({\hat C}_{\ell-1})^{1/2}({\hat C}_{\ell})^{1/2}
({\hat C}_{\ell+1})^{1/2}}
 \label{defJ}
\end{eqnarray}
These quantities are invariant under rotations
and parity; they are also dimensionless and therefore less
dependent upon the power spectrum. Indeed for a Gaussian
process it may be proved~\citep{conf} that it is this normalised
 -- and not the un-normalized -- 
bispectrum that is statistically
independent from the power spectrum. This matter will be 
very important later when we compare previously found
North/South power spectrum asymmetries with asymmetries found
in the normalised bispectrum.

\section{The WMAP bispectrum}\label{flat}
The WMAP mission \citep{wmap} produced full sky CMB maps
from ten differencing assemblies (DAs); four in the W band (94 GHz), 
two V band (61 GHz), two
Q band (41 GHz), one Ka band (33 GHz), and one K band (23 GHz).
The K and Ka bands are dominated by galactic emission and
therefore neglected for cosmological analysis.  We use 
the ``foreground cleaned'' maps\footnote{Publicly available from 
http://lambda.gsfc.nasa.gov},
where the Galactic foreground signal, consisting of synchrotron,
free-free, and dust emission, was removed using the 3-band,
5-parameter template fitting method described in \cite{Bennett:2003ca}.
All maps are rendered using the HEALPix\footnote
{http://www.eso.org/science/healpix/index.html} package \citep{healp},
with pixelization level nside=512.

\begin{figure}
\centerline{\psfig{file=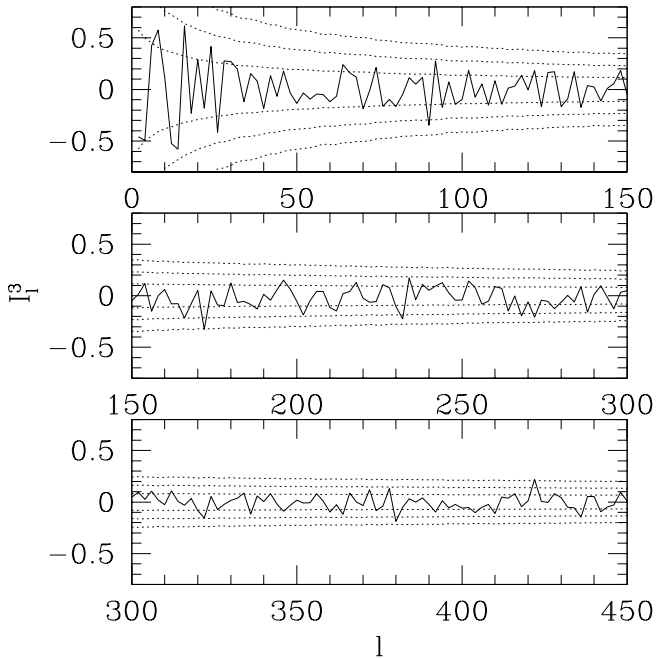,width=9cm}}
 \caption{The WMAP single-$\ell$ bispectrum for a coadded map
with the Kp0 mask.  We have superposed the 1, 2, and 3 sigma error
bars inferred from Gaussian simulations.}
\label{allI}
\end{figure}

\begin{figure}
\centerline{\psfig{file=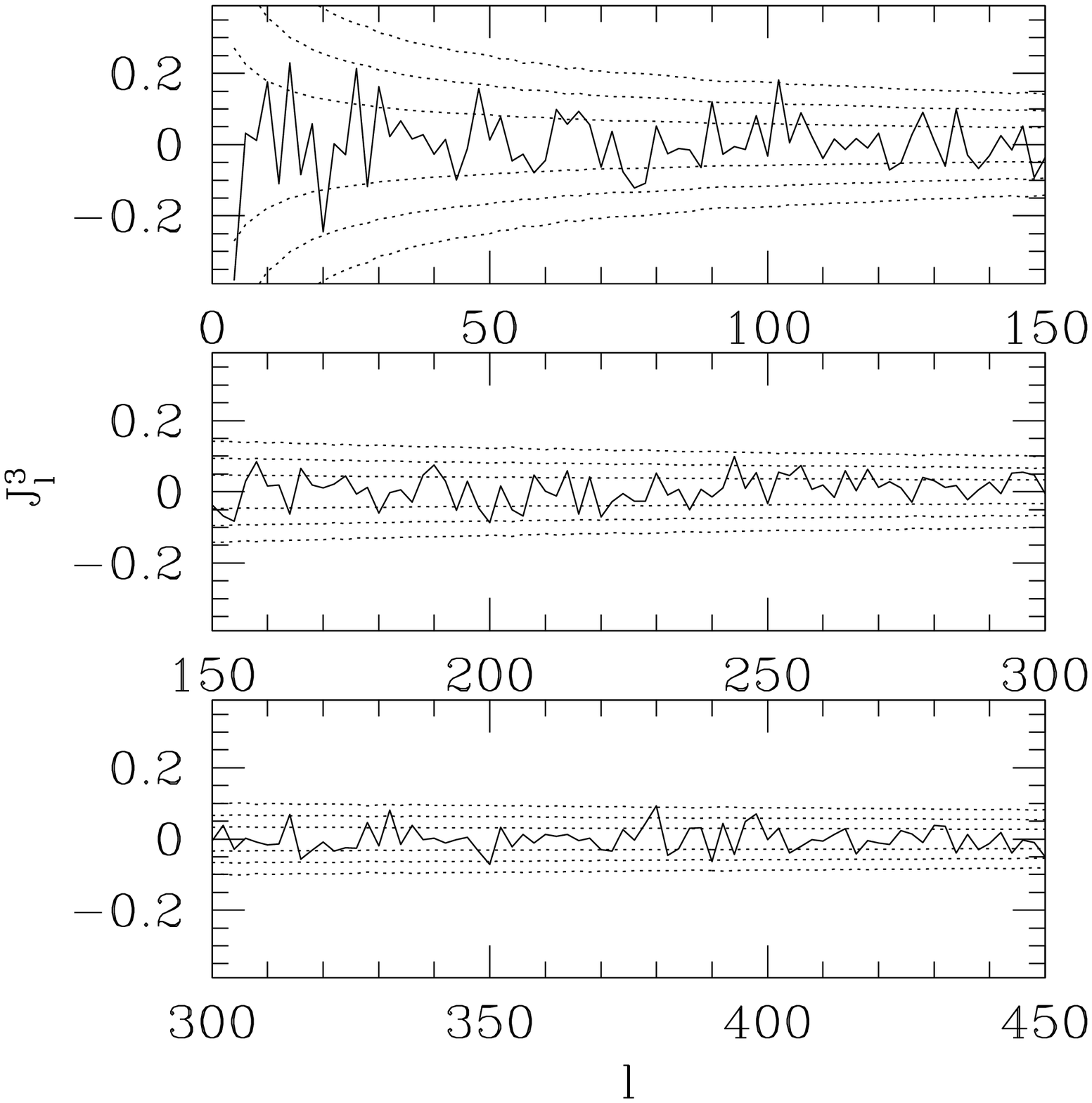,width=9cm}}
 \caption{The WMAP inter-$\ell$ bispectrum for a coadded map
with the Kp0 mask.  We have superposed the 1, 2, and 3 sigma error
bars inferred from Gaussian simulations.}
\label{allJ}
\end{figure}

We examined the bispectrum of an ``inverse-noise-squared'' coadded map 
based on the Q, V and W bands, that is a map:
\begin{equation}\label{coadd}
 T(n) =
\frac{\sum^{10}_{i=3} T_{i}(n) /
\sigma^{2}_{i}}{\sum^{10}_{i=3} 1 / 
\sigma^{2}_{i}}
\end{equation}
where
\begin{eqnarray}\nonumber
1 / \sigma^{2}_{i} = \sum^{Npix}_{n=1}{1 / \sigma^{2}_{i}(n)}
\hspace{2 em},\hspace{2 em}
\sigma^{2}_{i}(n) = \sigma^{2}_{0,i}/Nobs_{i}(n)
\end{eqnarray}
where $T_{i}$ is the sky map for the DA $i$ with the foreground
galactic signal subtracted, and $\sigma^{2}_{0,i}$ is the noise
per observation for DA i, whose values are given by \cite{wmap}. 
We apply the Kp0 mask to this map \citep{Bennett:2003ca} 
to cut possible remaining galactic contamination near the disk,
 and then we remove any residual monopole.

In Figs.~\ref{allI} and \ref{allJ} we plot the 
$I^3_\ell$ and $J^3_\ell$ bispectrum components, for $\ell=2-450$, and 
$\ell=4-450$ respectively,
for the WMAP first year coadded map
with a Kp0 mask. We compare these results with those from Monte
Carlo simulations of Gaussian maps with the same noise and beam
characteristics as the WMAP instrument, and subject to the Kp0 mask. 
From 5,000 sky realizations
we inferred the 1, 2, and 3-sigma variance lines, also 
plotted in Figs.~\ref{allI} and \ref{allJ}.

\begin{figure}
\centerline{\psfig{file=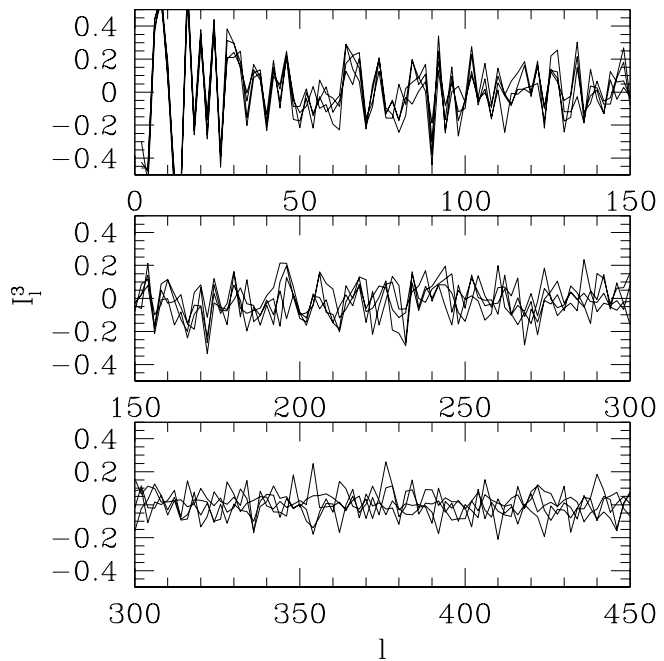,width=9cm}}
 \caption{The WMAP single-$\ell$ bispectrum for the channels that 
most contribute to the coadded map (Q1, Q2,
V1 and V2). }
\label{Ich}
\end{figure}

\begin{figure}
\centerline{\psfig{file=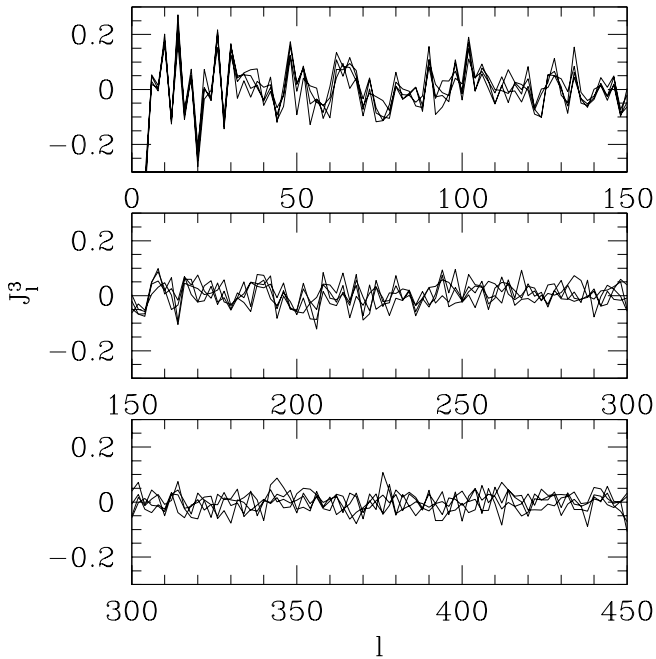,width=9cm}}
 \caption{The WMAP inter-$\ell$ bispectrum for the channels that 
most contribute to the coadded map (Q1, Q2,
V1 and V2). }
\label{Jch}
\end{figure}

Before we apply a goodness of fit statistic to these 
results it is important to select a signal dominated 
$\ell$  range. This should be estimated from {\it real} 
data; any other algorithms risk including noise-dominated
multipoles into the analysis, thus ending up testing the
Gaussianity of noise. Such a direct estimate may be obtained
by evaluating $I^3_\ell$ and $J^3_\ell$ 
for the different channels. The noise in each of these bands 
is uncorrelated, so the impact of the noise upon the statistics
may be inferred from the dispersion  from channel 
to channel. As can be seen in Figs.~\ref{Ich} and \ref{Jch}
the bispectrum is signal dominated up to $\ell\approx 200$.

\begin{table}
\centering
\caption{The fraction of our simulations that found a lower $\chi^2$}
 \begin{minipage}{30mm}

\begin{tabular}{lll} 
\hline
l range & $J^3_\ell$ & $I^3_\ell$ \\
\hline
2(4)-200 & 0.75 & 0.62 \\
2(4)-100 & 0.48 & 0.58 \\
2(4)-50   & 0.466 & 0.617 \\
30-70  & 0.239 & 0.033 \\
32-62  & 0.141 & 0.004 \\
50-100 & 0.409 & 0.437 \\
76-126 & 0.758 & 0.642 \\
100-150 & 0.854 & 0.445 \\
126-176 & 0.79  & 0.82 \\
150-200 & 0.746 & 0.617\\
\hline
\end{tabular}
\end{minipage}
\label{allchi}
\end{table}

Bearing this in mind we may now compute the reduced chi-squared
for a selection of signal dominated sections in $\ell$ space. 
Table~\ref{allchi} records the fraction of simulations that found a
lower $\chi^2$ value for the indicated $\ell$ range (note that the $\ell$ 
range can start at $\ell=4$ for $J_\ell$ and at $\ell=2$ for $I_\ell$). 
We find the WMAP
$J^3_\ell$ consistent with Gaussianity. However, for the $I^3_\ell$ we
observe a long ``flat patch'': a region where all points are within 1
sigma.
Our ``flat patch'' covers points $\ell=32-62$ inclusively, as can be 
seen in Fig.~\ref{allI}. For
this range we would expect a low $\chi^2$ value, and at 0.329 it is
smaller than 99.6\% of those from simulations for this
range. Critically we need to take account of selection effects, and so
by letting each simulation pick out its longest ``flat patch'' within
$\ell=2-300$ we can account for the fact that we have honed into the
most significant region. We find that a ``flat patch'' of this length or
longer appears in 8.8\% of the simulations. Further, a $\chi^2$ of 0.329 is
not at all significant for a ``flat patch'' the same length as ours:
67.7\% having an even lower $\chi^2$. Of course the last conclusion
depends on whether or not a prior exists which makes the range
$\ell=32-62$ special. If such a prior exists then we have rejected
Gaussianity at a high level of confidence. We shall return to this matter
in Section~\ref{select}.

It is important to evaluate the impact of the galactic cut on
these conclusions.
We note that the ILC map \citep{wmap} also displays this ``flat patch'', 
but only
when the central galactic area is masked. The ILC map does not completely
avoid foreground emissions, and so the galactic region will display
non-Gaussian signals. When the ILC map is not masked, it is this
galactic emission that dilutes our ``flat patch''. For both maps the
``flat patch'' does not diminish as we increase the masked region,
implying that the ``flat patch'' is not due to galactic emission.

We conclude that the the normalised inter-$\ell$ and single-$\ell$
bispectrum from the whole sky are consistent with Gaussianity.

\section{Hemisphere asymmetries in the bispectrum}\label{asym}

Recently there have been numerous detections of asymmetry in the CMB
(e.g., \cite{copi,coles,erik2,hansen,hbg,erik1,copi1}). 
Specifically, \cite{erik1} found a significant lack of power 
and a surprisingly featureless pseudo
collapsed 3-point correlation function, $C^{(3)}(\theta)$, in the
Northern ecliptic hemisphere. The asymmetry in the power spectrum was
maximised for a North pole at approximately 
$(\ell,b)=(57,10)$\footnote{Galactic coordinates: 
Longitude $\ell\in[-180(E),+180(W)]$, Latitude $b\in[-90(S),+90(N)]$, 
with $(\ell,b)=(0,0)$ towards the galactic centre.}.
The non-normalised bispectrum is related to the
Legendre transform of $C^{(3)}(\theta)$. Therefore by observing the
$I^3_\ell$ and $J^3_\ell$ on different hemispheres we can probe any
$\ell$ dependence of the $C^{(3)}(\theta)$ asymmetry. We emphasise
that these will also be independent of the power spectrum for a 
Gaussian field (see~\cite{conf} for an anlytical proof for $\ell=2$). 

\subsection{Inter-$\ell$ normalised bispectrum}\label{asyminter}

\begin{figure}
\centerline{\psfig{file=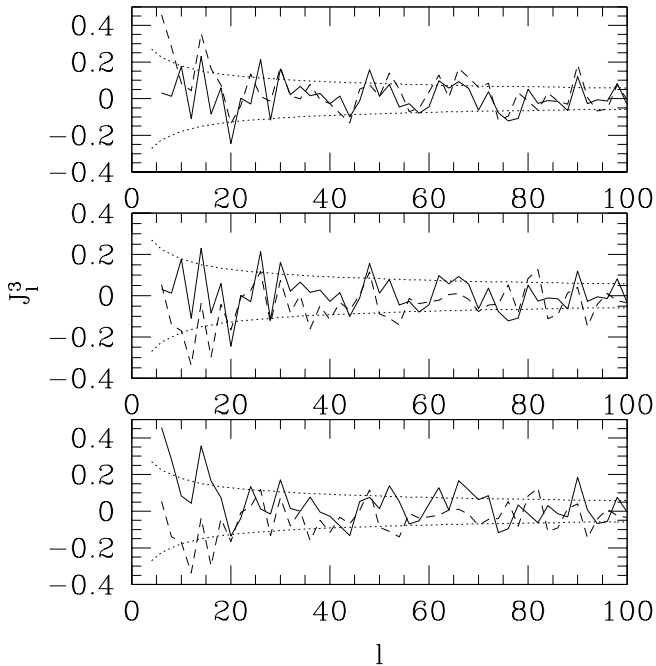,width=9cm}}
\caption{The $J^3_\ell$ spectra: top panel, allsky (bold) and
Northern (dashed); middle panel, allsky (bold) and Southern (dashed);
bottom panel, Northern (bold) and Southern (dashed).}
\label{N/S-J}
\end{figure}

We calculate the $J^3_\ell$ on the Northern and
Southern hemispheres for a North pole at $(\ell,b)=(57,10)$.
In Fig.~\ref{N/S-J} we plot them against each other, and 
against the whole sky results. We also plot the 1 sigma 
variance lines for the whole sky (the spread is a little wider for a half-sky).
We notice that the Northern results appear to take higher values than the
Southern. For 34 of the 49 points the Northern result is higher, and
when it is not the difference is small. We note that the 
individual $\chi^2$ values
do not reveal any significant departure from Gaussianity for either
the Northern or Southern results. It is the difference between the
hemispheres that is significant. 

We quantify this feature by measuring the
sum of the differences:
\begin{eqnarray}\label{K}
K &=& \sum_\ell\frac{J^3_{\ell,N}-J^3_{\ell,S}}{\sigma_\ell}
\end{eqnarray}
We calculate $K$ for a North pole at (57,10), 
and at the ecliptic North pole position (96,30), and at a further 53
 positions uniformly distributed 
in the Northern galactic hemisphere. We infer the value
of $K$ for the opposite North pole positions, in the Southern galactic
hemisphere.
We then perform Gaussian Monte Carlo simulations, as above, 
in order to evaluate the statistical significance of the
observed WMAP asymmetry. For each simulation we allow the North pole 
to vary over the same 
positions and we record the maximum $|K|$ value seen in each
simulation.

\begin{figure}
\centerline{\psfig{file=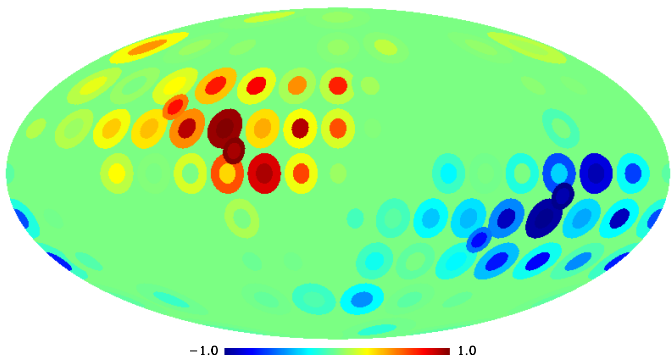,angle=90,width=9cm}}
 \caption{Asymmetry in the $J^3_\ell$ as measured by $|K|$. 
At each North pole position we show the fraction of simulations with a
lower $|K|_{\it max}$ value, multiplied by the $+/-$ sign of the 
$K_{\it WMAP}$ result. Results are also shown for North pole position 
(57,10) and its opposite (-123,-10), and for the ecliptic poles.} 
\label{K2}
\end{figure}

Figure~\ref{K2} plots, at the
corresponding North pole position, the fraction of simulations with a
lower $|K|_{\it max}$ value than the $|K_{\it WMAP}|$ value from that North
pole. This fraction is then displayed with the $+/-$ sign of this
$K_{\it WMAP}$ result {\it i.e.} -1, and +1 indicates a North pole
position returning higher asymmetry than all the simulations. We show
the results for the ranges $\ell=4-100$ (larger circles), and
$\ell=4-50$ (smaller circles).

In Table~\ref{Ktab} we list the results for the three
North pole positions that find the maximum value of $K$, for WMAP. 
We also show the fraction of simulations that return a lower
$|K|_{\it max}$ value.

\begin{table}
 \centering
\caption{Measuring asymmetry in the $J^3_{\ell}$ with $K$.  
The three North poles that maximise the asymmetry for WMAP, and
the fraction of simulations with a lower value of $|K|_{\it max}$.}
\begin{tabular}{ll} 
\hline
$\ell=4-150$&  \\
\hline
(57,10)&0.808\\
(63.9,20)&0.707\\
(85.1,20)&0.689\\
\hline
$\ell=4-100$&  \\
\hline
(57,10)&0.994\\
(63.9,20)&0.946\\
(40,0)&0.832\\
\hline
$\ell=4-50$& \\
\hline
(63.9,20)&0.988\\
(57,10)&0.922\\
(40,0)&0.916\\
\hline
\end{tabular}
\label{Ktab}
\end{table}

We observe significant asymmetry at the $\sim99\%$ level for both 
$\ell$ ranges 4-100, and 4-50. The North pole positions 
that maximise this asymmetry differ slightly between the 
$\ell$ ranges, but both are consistent with the 
(57,10) position that \cite{erik1} found maximises the angular power 
spectrum asymmetry. Furthermore, the pattern on the sky displayed in 
Fig.~\ref{K2} is very similar to that for the 
distribution of power, see Fig. 24 in \cite{hbg}. 

Whether this anomaly is a sign of anisotropy, non-Gaussianity
or both is partly rhetorical. An anisotropic ensemble 
may always be converted into an isotropic one by randomising
the preferred axis. From the point of view of such an ensemble,
what appears in a single sky as a sign of anisotropy is simply
non-Gaussianity (see~\cite{aniso} for details).

As previously mentioned, for a Gaussian random field the normalised 
bispectrum and the power spectrum are independent quantities. 
In the analysis in this section we were particularly careful 
to avoid selection effects (we let Gaussian 
simulations choose their maximal asymmetry axis). Should one
insist that the anomaly found is a fluke, however, we should
stress that it would be an independent fluke with respect to
the anomaly found in the power spectrum~\citep{erik1}.
This makes it even more unlikely. We shall return to this matter
in Section~\ref{select}.

\subsection{Single-$\ell$ normalised bispectrum}

\begin{figure}
\centerline{\psfig{file=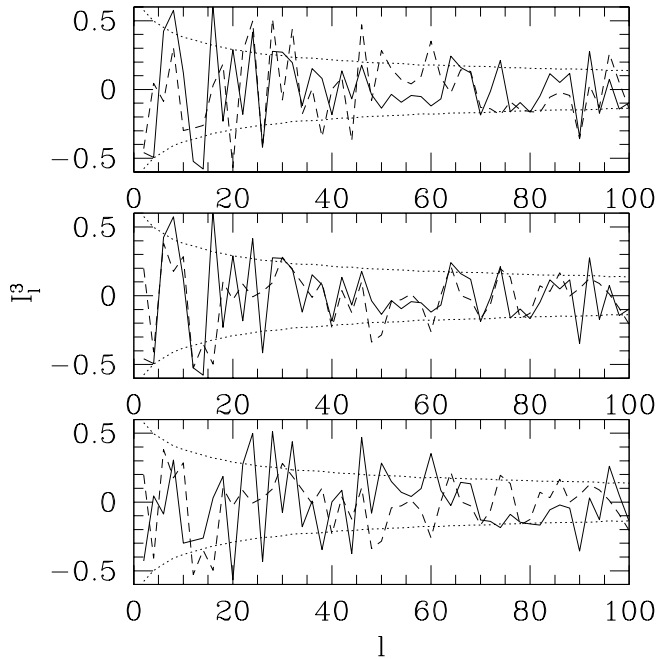,width=9cm}}
\caption{The $I^3_\ell$ spectra: top panel, allsky (bold) and
Northern (dashed); middle panel, allsky (bold) and Southern (dashed);
bottom panel, Northern (bold) and Southern (dashed).}
\label{N/S-I}
\end{figure}

We calculate the single-$\ell$ normalised bispectrum, $I^3_\ell$,
 on the Northern
and Southern hemispheres for a North pole at
$(\ell,b)=$(57,10). In Fig.~\ref{N/S-I} we plot them against each other, 
and against the whole sky results. We
also plot the 1 sigma error bars for the whole sky (the spread is very
similar for a half-sky).
The Northern results appear to fluctuate more 
widely than the Southern, and to investigate this further we calculate 
the $\chi^2$ values for both hemispheres, for various $\ell$ ranges.

In Table~\ref{NSchi} we show the fraction of simulations that find lower 
$\chi^2$ values, with our simulations also using a North pole at 
(57,10).
The WMAP Northern $\chi^2$ values are consistently high, while the 
Southern are low. For $\ell=$2-100 the Southern $\chi^2$ is low 
at the 99\% level. However, the significance of this asymmetry is not seen 
in these individual values from the North and the South, but in 
the ratio of the two: the fact that one is low 
{\it while} the other is high. Therefore, 
in Table~\ref{NSchi} we also record the fraction of simulations which find
a lower ratio, 
MAX($\frac{\chi^2_N}{\chi^2_S}$,$\frac{\chi^2_S}{\chi^2_N}$).

\begin{table}
 \centering
\caption{Measuring asymmetry in the $I^3_{\ell}$ with $\chi^2$. 
The fraction of simulations which have a lower $\chi^2$ for the South, 
a lower $\chi^2$ for the North, and a lower ratio of the two.}
\begin{tabular}{llll} \hline
l range & S & N & N/S \\ \hline
2-200 & 0.10 & 0.88 & 0.96\\
2-150 & 0.03 & 0.91 & 0.99\\
2-100 & 0.01 & 0.88 & 0.99\\
32-62 & 0.14 & 0.88 & 0.95\\
2-50 & 0.12 & 0.95 & 0.97\\
2-40 & 0.07 & 0.83 & 0.96\\
\hline
\end{tabular}
\label{NSchi}
\end{table}

\begin{figure}
\centerline{\psfig{file=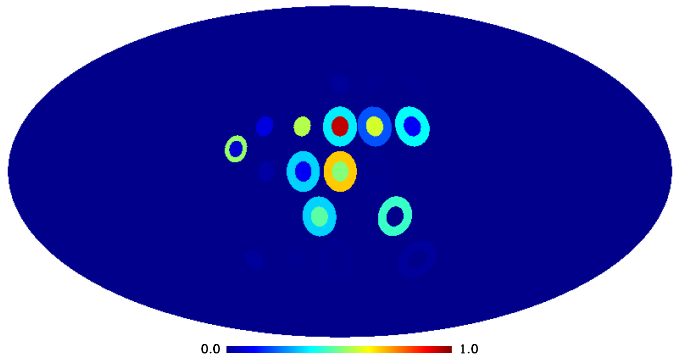,angle=90,width=9cm}}
 \caption{Asymmetry in the $I^3_\ell$ with the N-S $\chi^2$ ratio. 
Fraction of simulations with a lower maximum ratio.}
\label{mark-chi}
\end{figure}

Our $I^3_\ell$ asymmetry appears significant at the 99\% level for a North 
pole at (57,10). However, we must take account of our North pole 
selection as we may have pre-selected the North pole that maximises 
this asymmetry while 
we have not done the same for the simulations. As before, we vary the 
North pole position for the WMAP and for the simulations through a 
further 53 positions that are uniformly distributed 
in the Northern galactic hemisphere (and infer the results for the 
positions on the Southern galactic hemisphere). 
At every North pole position we calculate $\chi^2_N/\chi^2_S$, 
and $\chi^2_S/\chi^2_N$, and for each simulation we record the 
maximum chi-squared ratio result from all these values. 
In Table~\ref{NSchi1} we record 
the North pole positions which maximise the ratio for the 
WMAP data, and the fraction of simulations 
with lower maximum ratio values. We visualise this in Fig.~\ref{mark-chi}.

\begin{table}
 \centering
\caption{Measuring asymmetry in the $I^3_{\ell}$ with the North-South 
$\chi^2$ ratio. The fraction of simulations which, when varying the 
North pole, find a lower maximum ratio.}
\begin{tabular}{ll} 
\hline
$\ell=2-150$&  \\
\hline
(57,10)&0.581\\
(340.5,20)&0.180\\
(149.0,20)&0.108\\
\hline
$\ell=2-100$&  \\
\hline
(180,0)&0.683\\
(57,10)&0.527\\
(149,20)&0.431\\
\hline	
$\ell=2-50$& \\
\hline
(0,20)&0.940\\
(340.5,20)&0.587\\
(21.3,20)&0.557\\
\hline
\end{tabular}
\label{NSchi1}
\end{table}

The significance of our asymmetry detection in the single-$\ell$
bispectrum  has weakened greatly now that we have taken a selection
effect into  account. In fact the only hint of asymmetry that survives
is at  a level of 94\%, for the range $\ell=$2-50 with North pole at
(70,0). This result is not significant enough to rule out Gaussianity. 
In Fig.~\ref{mark-chi} we see that very few positions  have a
chi-squared ratio comparable to the simulations' maxima.  Further,
those that do, group together towards the galactic centre. This is a
very different pattern from that observed for the inter-$\ell$
bispectrum asymmetry, implying that any asymmetry here is unrelated.
The grouping towards the galactic centre does raise some concern that the
mask or foreground contamination may have a subtle effect on the 
single-$\ell$ bispectrum.

\subsection{3-point correlation function}

We briefly comment on the three-point correlation function, $C^{(3)}(\theta)$,
and
 its asymmetry reported by \cite{erik1}. We refer to this
work for a direct definition (in position space) of $C^{(3)}(\theta)$.
The $C^{(3)}(\theta)$, is related to our definition of the 
bispectrum ${\hat B_{\ell_1\ell_2\ell_3}}$, by
\begin{eqnarray}\label{3pt}
C^{(3)}(\theta)&=&\sum_{\ell_1\ell_2\ell_3}{\hat
B}_{\ell_1\ell_2\ell_3}P_{\ell_3}(cos(\theta))\times\\ &&{\left
(\begin{array}{ccc} \ell_1 & \ell_2 & \ell_3 \\ 0 & 0 & 0\end{array}
\right
)^{2}}\frac{(2\ell_1+1)(2\ell_2+1)(2\ell_3+1)}{(4\pi)^{\frac{3}{2}}}
\nonumber\\
&\equiv&\sum_{\ell_1\ell_2\ell_3}{{X}_{\ell_1\ell_2\ell_3}}
P_{\ell_3}(cos(\theta))\nonumber
\end{eqnarray}
It is computationally advantageous  to calculate $C^{(3)}(\theta)$ in
Fourier space -- through this summation -- than directly over all points in
the sky. Initially, we follow \cite{erik1} and choose not to include the 
dipole and quadrupole terms, because of their well 
known anomalous behaviour; therefore we perform the 
summation in equation (\ref{3pt}) for $\ell\geq$3. 
\begin{figure}
\centerline{\psfig{file=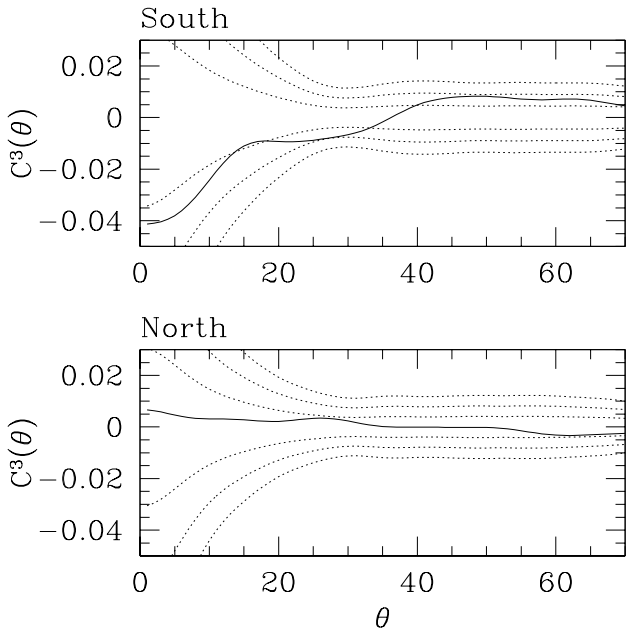,width=9cm}}
\caption{The threepoint correlation function, $C^{(3)}(\theta)$ in 
$10^6 (\mu K)^3$, from 
the Northern and Southern ecliptic hemispheres, with 1,2 and 3 sigma 
error bars from simulations.}
\label{plot3pt}
\end{figure}

In Fig.~\ref{plot3pt} we plot the results from the Northern and 
Southern \emph{ecliptic} hemispheres (scaled by their respective 
areas), with 1,2 and 3 sigma error bars 
from simulations. As can be appreciated from these
plots, we have 
confirmed the findings of \cite{erik1}: a featureless 
$C^{(3)}(\theta)$ in the Northern ecliptic hemisphere.
It is unlikely, however, that this result is simply related
to the bispectrum asymmetry reported in Section~\ref{asyminter},
as we shall now argue.

The three-point correlation function is a function in {\it real} 
space, while the bispectrum and power spectrum are in {\it Fourier} 
space. It is often not intuitively clear how a 
feature in one will manifest itself in the other. 
The $|{X}_{\ell_1\ell_2\ell_3}|$ values decrease very 
quickly with increasing $\ell$, so it is the earliest values that 
contribute the most, and determine the overall shape of 
$C^{(3)}(\theta)$. Including $\ell\geq$8 terms returns $C^{(3)}(\theta)$ 
values approximately an order of magnitude less than when 
including $\ell\geq$2 terms. Therefore, an effect that looks to propagate 
through a large range in real space, could be from a small 
number of values at this low-$\ell$ end in Fourier space. 
Taking this into account 
could decrease the significance of such a 
detection of asymmetry. In light of current uncertainties about 
foreground contamination in the low-$\ell$ multipoles 
\citep{slos, bgb, hbg} we investigate if the 
asymmetry survives when we exclude low-$\ell$ contributions.

We repeat the calculation of $C^{(3)}(\theta)$ 
on the Northern and Southern hemispheres, and each time we 
increase the minimum $\ell$ value allowed to contribute, {\it i.e.}, 
start the summation in equation (\ref{3pt}) from an increasing 
minimum $\ell$ value. We do the same 
for our simulations. For each computation, we perform a reduced 
chi-squared analysis over the $\theta$ range [0,70], and in 
Table~\ref{Tab3pt} we record the fraction of $\sim$500 simulations 
that find a lower $\chi^2$ value than the Northern and Southern 
WMAP results. Our simulations also use the ecliptic poles. 
We stress that we 
are not interested in the level of significance of this anomaly (we 
have not accounted for selection effects), but how the 
significance changes as we exclude low-$\ell$ terms.
We find the asymmetry does not diminish as we exclude 
the lower-$\ell$ contributions, implying an asymmetry that is a 
widespread feature in $\ell$ space.

\begin{table}
\centering
\caption{Excluding lower $\ell$ contributions in the calculation 
of $C^{(3)}(\theta)$. The fraction of simulations that found a 
lower $\chi^2$ from the Northern and Southern ecliptic hemispheres.}
 \begin{minipage}{30mm}
\begin{tabular}{lll} 
\hline
Min $\ell$ & $\chi^2_N$ & $\chi^2_S$ \\
\hline
2 & 0.066 & 0.716 \\
3 & 0.125 & 0.912 \\
4 & 0.068 & 0.521 \\
5   & 0.074 & 0.442 \\
6  & 0.000 & 0.601 \\
7  & 0.020 & 0.724 \\
8 & 0.009 & 0.687 \\
\hline
\end{tabular}
\end{minipage}
\label{Tab3pt}
\end{table}

We emphasise that this asymmetry is different to that 
explored above with $I^3_\ell$ and $J^3_\ell$, as here we are 
not independent of the power spectrum.

\section{The quadrupole}\label{quadtxt}
The geometrical meaning of the bispectrum becomes clearer by focusing on
the quadrupole.  Also, the low value of the WMAP quadrupole has inspired 
much speculation; our remarks shed some light on this matter.

In general a given multipole may be decomposed into the angular power
spectrum $C_\ell$ (the radius) and an axis (a direction), in $2\ell+1$
dimensions. Alternatively the latter may be represented as
$\ell$ independent unit vectors in 3 dimensions~\citep{copi,copi1}.
These contain the remaining $2\ell$ degrees of freedom (d.o.f);
but given that there are 3 rotational d.o.f. in the choice of the
coordinate frame (and that each multipole
is an irreducible representation of the rotation group),
such vectors cannot be rotationally invariant, i.e. they depend on the way the
system of coordinates was set up. 

For instance the quadrupole has a total of 5 d.o.f. One is $C_2$,
so there is only one other invariant. The 2 3D axes resulting from
the procedure in~\cite{copi} contain all the remaining 4 degrees of
freedom, so they mix
the issues of non-Gaussianity and anisotropy.
In fact there are three independent aspects to quadrupole anomalies: 
the low value of 
the angular power spectrum ($C_2$), its possible non-Gaussianity,
and its possible anisotropy. 
\begin{table*}\label{quad}
 \centering
 \begin{minipage}{70mm}
  \caption{The Quadrupole bispectrum and eigenvectors; and
some ``unusual directions''.}
  \begin{tabular}{@{}l|r|rr|rr|rr@{}}
  \hline

Dataset & $I^3_2$   &  \multicolumn{2}{c}{${\bf v}_1$} &
\multicolumn{2}{c}{${\bf v}_2$} &\multicolumn{2}{c}{${\bf v}_3$} \\
&  & $l$ & $b$ & $l$ & $b$ & $l$ & $b$ \\
\hline
Coadd, Kp0 & -.458 & 48.01 &69.39 &-118.43&20.08&149.94&4.48\\
ILC, Kp0 &-.439 &41.26&63.49&-115.40&24.60&150.32&9.25\\
ILC & -0.386 &-82.66&68.56&66.02&18.54&-20.46&-10.38\\
Kp0 & 0.771 &-159.78&83.95&172.41&-5.35&82.67&2.80\\
\hline
Dipole & --- & -96.15 & 48.25 & -- & -- & -- &-- \\
CTH (Virgo) & --- & -110. & 60. & -- & -- & -- &-- \\
${\bf w}^{(2,1,2)}$& --- & -105.73 & 56.62& -- & -- & -- &-- \\
\hline
\end{tabular}
\end{minipage}
\end{table*}

An alternative approach is that pioneered in~\cite{conf} (the
two approaches actually coincide for $\ell=1$). By rewriting 
the spherical harmonics in Cartesian coordinates (instead of
their more popular rendition in polar coordinates) one obtains a symmetric
multilinear form of order $\ell$. Thus the $\ell$ multipole may be regarded as 
a traceless, symmetric, rank $\ell$ tensor. From this one may extract
$2\ell-2$ (invariant) contractions, and a orthonormal frame. For example
the quadrupole may be regarded as a $3\times 3$ symmetric traceless
matrix. By considering Cartesians $x^i=(z,x,y)$ (with constraint 
$x^2+y^2+z^2=1$) one has:
\bea
Y^2_0&=&\sqrt {5\over 16 \pi}(2z^2-x^2-y^2)\\
Y^2_{\pm 1}&=&\mp\sqrt {15\over 8 \pi}(x\pm i y)z\\
Y^2_{\pm 2}&=&\sqrt {15\over 32 \pi}(x\pm i y)^2
\eea
Thus the quadrupole may be rewritten as
\be
\delta T_2=\sum_m a_{2m} Y_{2m}=Q_{ij}x^i x^j
\ee
leading to matrix
\be
Q_{ij}=\sqrt{5\over 16 \pi}
{\left( \begin{array}{ccc} {2 a_{20}\over {\sqrt 3}}
 & -{\sqrt 2} a_{21}^{re} & {\sqrt 2} a_{21}^{im} \\ 
-{\sqrt 2} a_{21}^{re} & - {a_{20}\over {\sqrt 3}}+ {\sqrt 2} a_{22}^{re} & 
-{\sqrt 2} a_{22}^{im}\\
{\sqrt 2} a_{21}^{im}& -{\sqrt 2} a_{22}^{im} & - {a_{20}\over {\sqrt 3}}- {\sqrt 2} a_{22}^{re} 
\end{array}\right)}
\ee
The 5 d.o.f. in this matrix split into 
2 invariants and the 3 d.o.f. associated with coordinate axes' orientation;
for example  2 independent eigenvalues (the 3
eigenvalues must add to zero) and 3 orthonormal eigenvectors.
The two invariants may also be the square of the matrix
(the sum of the squares of its eigenvalues) and its determinant
(the product of the eigenvalues). The first invariant is the power spectrum,
the latter the bispectrum. However for a Gaussian
these are not statistically independent, since in 3D
they must satisfy the constraint
\be
|det Q|\le {Q^{3}\over 3\sqrt{6}}
\ee
Thus, one should consider the normalised bispectrum, here
rewritten as:
\be
I^3_2={3\sqrt{6}\det{Q}\over Q^{3}}
\ee
if we want to separate power spectrum effects from non-Gaussianity
issues. The variables $C_2$, $I^3_2$, and the vector system are
now independent random variables; $C_2$ has a $\chi^2_5$ distribution,
$I^3_2$ a uniform distribution in $[-1,1]$, and the eigenvectors
are uniformly distributed. 

The issues of isotropy and non-Gaussianity are now fully separated. 
We have checked with Monte Carlo simulations that indeed $I^3_2$
and the eigenvectors are uniformly distributed when considering
a Gaussian as seen through WMAP.
Do the quadrupole eigenvectors point in 
a suspicious direction? We plot results in Table~\ref{quad}, for the 
inverse noise squared coadded map, with Kp0 mask, for the internal linear
combination map with and without the Kp0 mask, and for the mask alone.
None of the eigenvectors appears to correlate with the kinematic 
dipole direction, with the directions spotted by~\cite{copi1}, 
or with those found by~\cite{teg} (which is roughly in the direction
of the Virgo cluster).

We conclude that the bispectrum analysis provided in this paper
does not show anything abnormal for the quadrupole.

\section{Discussion: selection effects
and ``anomalies''}\label{select}
In the bispectrum studies above we were very conservative
about selection effects. For instance, the flat patch found in
Section~\ref{flat} was dismissed on the grounds that any Gaussian
realization will reveal a flat patch {\it somewhere} if a sufficiently
long string of $\ell$s is considered (a similar argument
did not degrade the detection in Section~\ref{asym}).
We stress that the flat patch effect may be
real if we take a closer look at selection effects, the
meaning of anomalies and the role of priors.

``Anomaly'': the very name conjures images of prejudice. 
And yet, it's extremely difficult to make statistical statements
without an a priori expectation. This expectation defines the 
``non-anomalous'' model. But if the model makes a large number of 
statistical predictions, looking into a sufficient number of 
observations will eventually turn out an anomaly even if the model
is right. Using this argument we rejected the flat patch 
described in Section~\ref{flat}.
However, it could also be that the anomaly is real, and is in fact
a vindication of an alternative model. In this case
our choice of prior and subsequent classification of the anomaly
as a fluke has dramatically prejudiced the analysis.

An interesting historical example is
the flat patch found in the COBE-DMR four year data~\citep{joao}. 
Taken on its own it could not be dismissed as a selection
effect (it covered most of the range to which the experiment
was sensitive). If WMAP had confirmed a flat patch in the
COBE range ($\ell=4-20$) but not elsewhere, then we would have the strange 
result that {\it more} data would {\it decrease} the significance of 
a detection. This obviously means that we have excessively favoured 
a Gaussian prior. 

More interesting is what actually happened: WMAP failed
to confirm the COBE flat patch. This  showed that indeed 
the observed COBE flat patch was {\it not}
a fluke: it was a systematic. \cite{jooes} explained in detail
the origin of the systematic error.

The rejection of the flat patch described in Section~\ref{flat} as 
a selection effect could be 
a further example of the dangers of blind use of 
a Gaussian prior. 
Let us consider a simplified model in which $N$ is the number
of observed independent bispectrum components and $p$ be 
the probability that each one of them is within $n$ sigma of the mean.
If we define a flat patch with $n=1$ we have $p\approx 0.68$.
The probability of a flat patch longer than $L$ starting at a given
fixed multipole is $p^L$. In WMAP a flat patch occurs with length $L=16$,
starting at $\ell=32$. Its probability for a Gaussian 
is $p_L=0.002$, seemingly ruling out Gaussianity. 

In Section~\ref{flat} we argued that a selection effect was at play. 
Using our simplified model, the 
probability of a Gaussian generating a flat patch longer than 
$L$ starting {\it anywhere} is
$p^L(N-L+1)$. In this case, considering that $N\approx 100$
(number of signal dominated bispectrum components), we get
a probability $0.18$, i.e. consistency with Gaussianity.
In Section~\ref{flat} we refined this argument with Monte
Carlo simulations.

The problem with this argument is that it is reversed if
there is a theory predicting a flat patch in the range
$\ell=32-62$, and this is used as a prior. Then 
Gaussianity is ruled out at the $99.6\%$ confidence level. So the prior 
is all important. 
As the example of COBE-DMR shows the threat of systematics
should make us beware of simply dismissing
anomalies as flukes. 

We stress that none of these remarks apply to the results in
Section~\ref{asyminter},  concerning North-South asymmetries in the
normalised inter-$\ell$ bispectrum. We were particularly careful to
compare like with like - letting the Gaussian simulations choose the
axis of maximal symmetry, just as we did with the data. 
The fact that the maximal asymmetry axis correlates
well with that found for the power spectrum asymmetry reinforces the
detection. For a Gaussian process power spectrum and normalised
bispectrum are independent random variables: one would need
two independent flukes instead of one.

\section{Summary}\label{con}

In this paper we have performed tests of Gaussianity and explored 
possible asymmetries of the WMAP first year data.
We calculated the inter-$\ell$, $J^3_\ell$, and single-$\ell$,
 $I^3_\ell$, normalised bispectrum and compared the 
WMAP results with those from simulations.
We found the whole sky results to be consistent with Gaussianity, 
barring a 'flat patch' anomaly in the $I^3_\ell$ for $\ell=32-62$. 
Depending on our attitude toward priors this may or may not
be seen as rejecting Gaussianity at the $99.6\%$ confidence level. 

We also searched for hemisphere asymmetries in the bispectrum,
and using the $J^3_\ell$ found a North-South asymmetry, with
significance  level maximised at 99\% for a North pole at 
$(\ell,b)\approx$(57,10). This 
is the same North pole that maximises angular-power spectrum 
asymmetry \citep{erik1,hbg}.
A similar analysis with $I^3_\ell$ showed consistency with Gaussianity.

We noted that asymmetries in the 3-point correlation function, $C^3(\theta)$, 
\citep{erik1} do not diminish when one excludes 
low-$\ell$ contributions. However such asymmetries are unlikely
to be related to those reported in this paper.
Our quadrupole analysis failed to find any
correlation with previously reported anomalies~\citep{copi1,teg}.

Whether these features are real or the hallmark of further
systematic errors remains to be seen.

\section*{Acknowledgements}
We would like to thank A.J. Banday, M. Kuntz and J. Medeiros 
for help with this project.
The results in this paper have been derived using the 
HEALPix package~\citep{healp}.

\bsp

\label{lastpage}

\end{document}